\documentclass{article}

\PassOptionsToPackage{numbers, compress}{natbib}

 \usepackage[main, final]{neurips_2026}

\usepackage[utf8]{inputenc} 
\usepackage[T1]{fontenc}    
\usepackage{hyperref}       
\usepackage{url}            
\usepackage{booktabs}       
\usepackage{amsfonts}       
\usepackage{amsmath}  
\usepackage{nicefrac}       
\usepackage{microtype}      
\usepackage{xcolor}         

\usepackage[numbers]{natbib}

\usepackage{comment}
\usepackage{colortbl}
\usepackage{multirow}
\usepackage{graphicx}
\usepackage{wrapfig}
\usepackage{ulem}
\usepackage[most]{tcolorbox}
\usepackage{float}
\definecolor{shadecolor}{RGB}{0, 125, 67} 

\definecolor{failcolor}{RGB}{240, 120, 0} 
\definecolor{SKKU_green}{RGB}{47, 79, 64}
\definecolor{Edit}{RGB}{30, 6, 209}

\definecolor{TODO}{RGB}{201, 52, 235}%

\newcounter{takeawayonly}

\newcommand{\parag}[1]{\vspace{+0.0mm}\noindent\textbf{#1}}
\newcommand{\takeawayonly}[1]{
    \vspace{-0.05cm}
    \refstepcounter{takeawayonly}
    \begin{tcolorbox}[
        colback=SKKU_green!8,                       
        colframe=SKKU_green!95,                     
        arc=4pt,                    
        boxsep=5pt,                 
        left=2pt,                  
        right=2pt,                 
        top=4pt,                    
        bottom=4pt,                 
        boxrule=0.8pt,              
        drop shadow=gray!30!white,  
        enhanced jigsaw             
    ]
    \vspace{-0.15cm}
        \parag{\textbf{}} #1
    \vspace{-0.15cm}
    \end{tcolorbox}
}

\title{Continual Speaker Identity Unlearning\\ with Minimal Interference}

\author{%
  Jinju Kim\textsuperscript{1} \quad
  Yunsung Kang\textsuperscript{1} \quad
  Gyeong-Moon Park\textsuperscript{2} \quad
  Jong Hwan Ko\textsuperscript{1} \\
  \textsuperscript{1}Sungkyunkwan University \quad
  \textsuperscript{2}Korea University \\
  \texttt{\{perla0328, jhko\}@skku.edu}
}

\begin{document}

\maketitle

\begin{abstract}
  Machine unlearning removes designated concepts or knowledge from pre-trained models. Recent work has extended this paradigm to speaker identity unlearning in zero-shot text-to-speech (ZS-TTS), the task of selectively erasing a model's ability to replicate a speaker's voice. Existing methods, however, quietly assume all unlearning requests arrive at once; an unrealistic assumption, since privacy-motivated removals arrive sequentially over time. We show this assumption breaks state-of-the-art methods: unlearning each new speaker fully revives previously unlearned speakers, reintroducing the very privacy risk unlearning was meant to eliminate. We present \textbf{C}umulative \textbf{ORT}hogonal \textbf{I}dentity \textbf{S}uppression (CORTIS), the first framework for continual speaker identity unlearning in ZS-TTS that requires no access to previously-unlearned speaker data. CORTIS combines Fisher-information-based parameter masking, which localizes updates to speaker-relevant weights, with orthogonal projection against subspaces spanned by prior unlearning updates. 
  With VoiceBox, CORTIS unlearns each requested speaker while keeping previously unlearned speakers forgotten across long request sequences, substantially outperforming sequential application of prior methods. The demo is available at \url{https://cumulativeortis.github.io/}.
\end{abstract}

\setcounter{tocdepth}{1}
\addtocontents{toc}{\protect\setcounter{tocdepth}{-1}}
\section{Introduction}
Imagine your family receiving a phone call in your voice saying something you would never say, and never said.
Modern Zero-Shot Text-to-Speech (ZS-TTS) systems can now clone a target speaker's voice from a reference utterance of only a few seconds \citep{le2023voicebox, valle, tan2024naturalspeech}. This capability, while technically impressive, has outpaced the safeguards around it: a short clip from a phone call, a podcast, or a voicemail is sufficient to synthesize convincing speech attributed to an individual who never produced it. In response, data-protection regulations such as the GDPR~\citep{gdpr} and the CCPA~\citep{ccpa} grant users the right to request removal of personal data from deployed systems, or the right to be forgotten (RTBF)~\citep{rtbf}. Translating this into a concrete procedure for deployed ZS-TTS systems is the problem we study. 

Machine unlearning \citep{bourtoule2021machine} provides the algorithmic framing. Recent work on speaker identity unlearning in ZS-TTS \citep{donotmimic} has shown that a model can be induced to produce random or non-identifying output when prompted with a forget-speaker reference, typically by distilling a randomly-initialized or noise-generating teacher on the forget set while preserving a retain loss. 
The setting assumed in prior, however, does not match deployment. 
Prior work assumes that all unlearning requests are known in advance and handled in a single simultaneous unlearning, with forgetting evaluated only against a fixed, persistently maintained forget set.

In reality, unlearning requests cannot arrive all at once: one user today, while another next month. 
Under this realistic setting, prior speaker identity unlearning method quickly falls into data retention paradox. 
To comply with RTBF and successfully unlearn, the provider must keep every requester's data for a simultaneous unlearning process in the future. 
Paradoxically, keeping every requester's data directly contradicts with the right being exercised, RTBF.
The natural alternative is to process each request on arrival then discard the data, leaving the unlearned weights safeguarded under release.

\begin{figure}
    \centering
    \includegraphics[width=\linewidth]{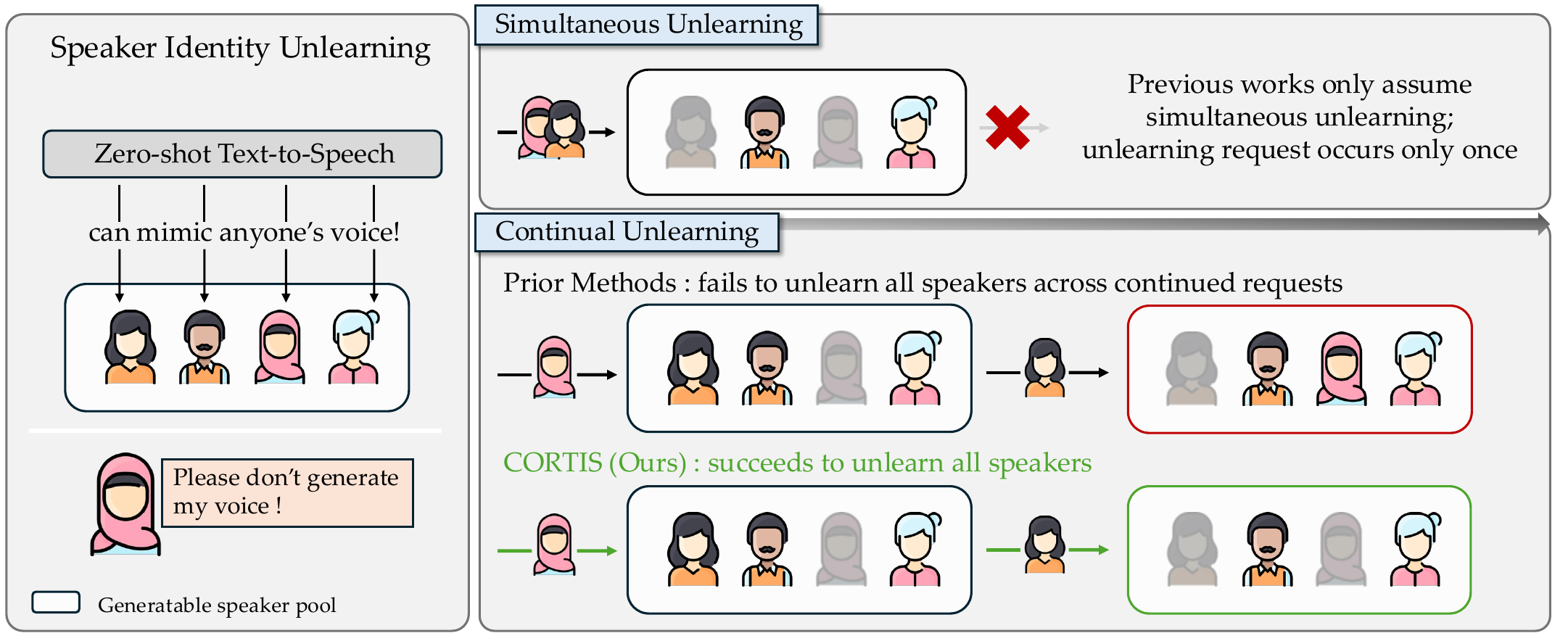}
    \caption{Na{\"i}vely applying prior speaker identity unlearning methods in continual sequence leads to failure. Ideally, previously unlearned speakers should remain suppressed across subsequent requests.}
    \label{fig1_task}
\end{figure}

However, as depicted in Figure \ref{fig1_task}, applying existing methods in this continual fashion fails, and fails in a way that matters for privacy: unlearning a new speaker causes measurable recovery of speakers that were already unlearned. 
We refer to this failure mode as \textit{catastrophic re-learning}, and trace it to a specific cause. 
The retain loss that protects model utility on subsequent requests provides no supervision on the parameters previously responsible for forgetting earlier speakers, leaving those parameters free to drift back. To address this, we propose \textsc{CORTIS}, a \textbf{C}umulative \textbf{ORT}hogonal \textbf{I}dentity \textbf{S}uppression. CORTIS specifically targets continual speaker identity unlearning with no access to previously-unlearned speakers' data.
We combine two complementary mechanisms. 
First, contrastive Fisher-information saliency localizes each unlearning update to a small set of parameters most relevant to forgetting the current speaker, while soft-excluding parameters important for prior speakers or for retain quality. 
Second, inspired by continual learning~\citep{farajtabar2020ogd, saha2021gradient}, we apply orthogonal projection to constrain the direction of the update. Specifically, we project updates onto the orthogonal complement of the cumulative subspace spanned by prior unlearning sequences, so that updates within the localized region cannot drift along directions previously used to forget earlier speakers.
On VoiceBox~\cite{le2023voicebox}, \textsc{CORTIS} keeps every forget speaker's similarity below 0.18 after five sequential unlearning requests, a 75\% average reduction from the pretrained baseline, while preserving competitive remain-set quality.

To summarize, our main contributions are as follows:
\begin{itemize}
    \item This is the first paper to formalize continual speaker identity unlearning for ZS-TTS under realistic deployment constraints; sequential request arrival and forget-data non-retention. We show that prior methods fail and identify catastrophic re-learning as the central failure mode.
    \item We propose \textsc{CORTIS}, a continual unlearning framework that combines contrastive Fisher-information parameter localization with cumulative orthogonal subspace projection. To our knowledge, this is the first method designed for continual speaker identity unlearning that does not require access to previously-unlearned speakers' data.
    \item We empirically demonstrate that \textsc{CORTIS} suppresses every previously forgotten speakers across sequential requests while retaining competitive zero-shot TTS quality on remain speakers, where every prior method either reverts forgotten identities or collapses utility.
\end{itemize}

\section{Related Works}


\subsection{Machine Unlearning}


Machine unlearning removes the influence of a subset of data or concepts from a trained model, motivated by regulations such as the GDPR Right to Be Forgotten~\citep{gdpr, rtbf}. 
Most existing work assumes a setting in which the entire forget set is specified up front and removed in one optimization pass~\citep{gandikota2023erasing, fan2024salun, heng2023selective, donotmimic}. 
In practice, however, forget requests arrive sequentially over a deployed model's lifetime~\citep{cooper2024machine}. Naively re-applying the unlearning method for each sequential request repeatedly diminishes model capabilities~\citep{liu2025rethinking}. This sequential setting, though only recently studied, is termed continual unlearning~\citep{jang2023knowledge}. 
Recent work has begun to address continual unlearning, primarily in computer vision~\citep{lee2025continual, george2025distill} and large language models~\citep{gao2025o3, xu2026fit, wuerkaixi2025alkn}, via mechanisms including gradient projection, distillation, orthogonally constrained adapters, and targeted parameter updates.

Speaker identity unlearning has only been examined in a limited setting. The first paper introduced the task for ZS-TTS and proposed Teacher-Guided Unlearning (TGU), which updates the model to produce random voices when prompted with forget speakers~\citep{donotmimic}. 
While effective at unlearning speakers jointly, TGU is evaluated only under a single training phase, where the full forget set is known in advance. 
The continual case  where requests arrive sequentially remains unaddressed.

\subsection{Continual Learning}



Continual learning addresses the setting in which a model must learn a sequence of tasks that arrive one at a time, and must maintain performance on all previously seen tasks without retraining from scratch~\citep{thrun1998lifelong, parisi2019continual}.
Continual learning requires that updates for task $t$ preserve the model utility and performance for tasks $1$, \ldots, $t-1$. 
The central obstacle is \textbf{catastrophic forgetting}~\citep{mccloskey1989catastrophic, kirkpatrick2017overcoming}: gradient steps taken to minimize loss on a new task (e.g., `washing dishes') overwrite parameters important for previous tasks (e.g., `folding laundry'), causing sharp drops in earlier-task performance despite the network having the capacity to solve all tasks jointly.

Existing methods fall into a variation or combination of four approaches. Regularization-based approaches discourage movement of parameters deemed important for prior tasks~\citep{kirkpatrick2017overcoming, schwarz2018progress, aljundi2018memory, zenke2017continual}.
A representative work EWC~\citep{kirkpatrick2017overcoming} uses the Fisher Information diagonal to weight a quadratic penalty, and hard-masking variants with important Fisher-ranked parameters. 
Replay-based approaches try to remember the past rather explicitly, using stored or generatively reconstructed samples from past tasks~\citep{rolnick2019experience, shin2017continual}. Parameter-isolation approaches minimize interference structurally by allocating disjoint subsets of parameters to each task~\citep{rusu2016progressive, mallya2018packnet, serra2018overcoming}. Optimization-based approaches constrain updates directly in gradient space rather than through the loss. This often involves projecting new-task gradients onto subspaces orthogonal to those of prior tasks~\citep{farajtabar2020ogd, saha2021gradient, zeng2019continual}; unlike regularizers, which permit drift into forbidden directions when the loss signal is strong enough, projection strictly removes such components regardless of loss magnitude.

\section{Background}\label{sec:motivation}

Our goal is to establish a realistic baseline for continual speaker identity unlearning in deployed ZS-TTS systems, where forget requests arrive sequentially over the model's lifetime rather than at once. In Sec.~\ref{subsec:problem}, we formalize the objectives of continual speaker identity unlearning, extending the one-shot formulation of prior work~\cite{donotmimic} to a stream of forget requests and articulating practical constraints such as the data retention paradox that distinguish the continual setting from its one-shot counterpart. In Sec.~\ref{subsec:motivation}, we argue that \textit{continual} speaker identity unlearning is not a straightforward extension from prior speaker identity unlearning but a fundamentally distinct task. We identify catastrophic re-learning of previously unlearned speakers as the central failure mode.

\subsection{Problem Formulation}\label{subsec:problem}

Let $\mathcal{S}$ denote the set of all generatable speakers, and let $\mathcal{D}^s = \{(x^s, y)\}$ denote a dataset of transcribed utterances for speaker $s \in \mathcal{S}$, where $x^s$ is an audio prompt and $y$ is its transcription. 
A pre-trained zero-shot TTS model $\theta_0$ satisfies $\theta_0(x^s, y) \approx \hat{x}^{spk=s}_{y}$ for any $s \in \mathcal{S}$, including speakers unseen during training, where $\hat{x}^{spk=s}_{y}$ denotes speech delivering text $y$ in the voice of $s$. We study continual speaker identity unlearning under two constraints that reflect realistic Right-To-Be-Forgotten (RTBF) deployment.

\textbf{(C1) Sequential arrival.}
Unlearning requests arrive one at a time. At step $i$, a request to forget a set of speakers $f_i \subseteq \mathcal{S} \setminus \mathcal{F}_{i-1}$, where $|f_i| \geq 1$, produces a cumulative forget set $\mathcal{F}_i = \mathcal{F}_{i-1} \cup f_i$ with $\mathcal{F}_0 = \emptyset$, and a corresponding remain set $\mathcal{R}_i = \mathcal{S} \setminus \mathcal{F}_i$. The system updates $\theta_{i-1} \rightarrow \theta_i$ without knowledge of future requests $f_{i+1}, f_{i+2}, \cdots$.

\textbf{(C2) Forget non-retention.}
Once step $i$ has been processed, $\mathcal{D}^{f_i}$ is discarded and cannot be accessed at any subsequent step $i' > i$. This excludes replay-based methods that require access and maintenance of prior forget sets $\mathcal{D}^{\mathcal{F}_{i-1}}$. 
The current model state $\theta_i$ satisfying the unlearn request and the remain set $\mathcal{D}^{\mathcal{R}_i}$ can persist across steps, whereas prior checkpoints and the forget data they were trained against must be discarded.



\textbf{Objective.}
At each unlearn sequence $i$, the unlearned model $\theta_i$ must satisfy the following conditions:
\begin{equation}
\theta_i(x^r, y) \approx \hat{x}^{spk=r}_{y} \quad \forall r \in \mathcal{R}_i, ~~\text{and}~~
\theta_i(x^f, y) \not\approx \hat{x}^{spk=f}_{y} \quad \forall f \in \mathcal{F}_i,
\end{equation}
preserving zero-shot synthesis quality for all current remain speakers while refusing to replicate any speaker in the cumulative forget set $\mathcal{F}_i$, not solely the most recent request $f_i$. The central technical challenge under (C1) and (C2) is to maintain the second condition for $\mathcal{F}_{i-1}$ while 
unlearning $f_i$, without access to $\mathcal{D}^{\mathcal{F}_{i-1}}$.

\subsection{Catastrophic Re-learning} \label{subsec:motivation}

Continual unlearning typically suffers from catastrophic forgetting or rapid utility collapse, in which the model loses overall capability as new unlearning requests are introduced. To prevent this, strong regularization is the standard remedy across domains \citep{lee2025continual}. For continual unlearning in ZS-TTS systems, catastrophic forgetting manifests in a distinctive way. As the model is inherently generalizable to unseen speakers, regularization that protects retain-set utility without an active forget loss on previously unlearned speaker is sufficient to revive that speaker during a subsequent unlearning sequence. We refer to this failure mode as catastrophic re-learning: the standard remedy for catastrophic forgetting in other domains becomes the very mechanism that undoes prior unlearning.

One might attempt to address catastrophic re-learning by re-introducing an active forget loss on every previously unlearned speaker at each new request. 
This appears straightforward, but it leads to two compounding problems. 
First, this is not meaningful continual unlearning anymore. The whole point of the continual setting is to make small, incremental updates rather than train from scratch. Re-introducing a forget loss on every prior speaker requires retaining all prior forget data, so both the data and the training cost grow with each step (\textit{e.g.}, $1K \rightarrow 2K \rightarrow 3K \rightarrow \cdots$ training steps as requests accumulate).
Second, and more fundamentally, retaining a forget loss on previously unlearned speakers requires retaining their data. This directly violates the deletion guarantee that motivated unlearning in the first place. Data retention paradox occurs: any method that satisfies durability through retained data fails RTBF compliance, and any method that satisfies RTBF by deleting the data loses the anchor it needs for durability.

\takeawayonly{Speaker identity unlearning demands a fundamentally different treatment of continual unlearning — suppress \textbf{catastrophic re-learning} on previously unlearned speaker identities.}

\section{Method}

\begin{figure}
    \centering
    \includegraphics[width=1\linewidth]{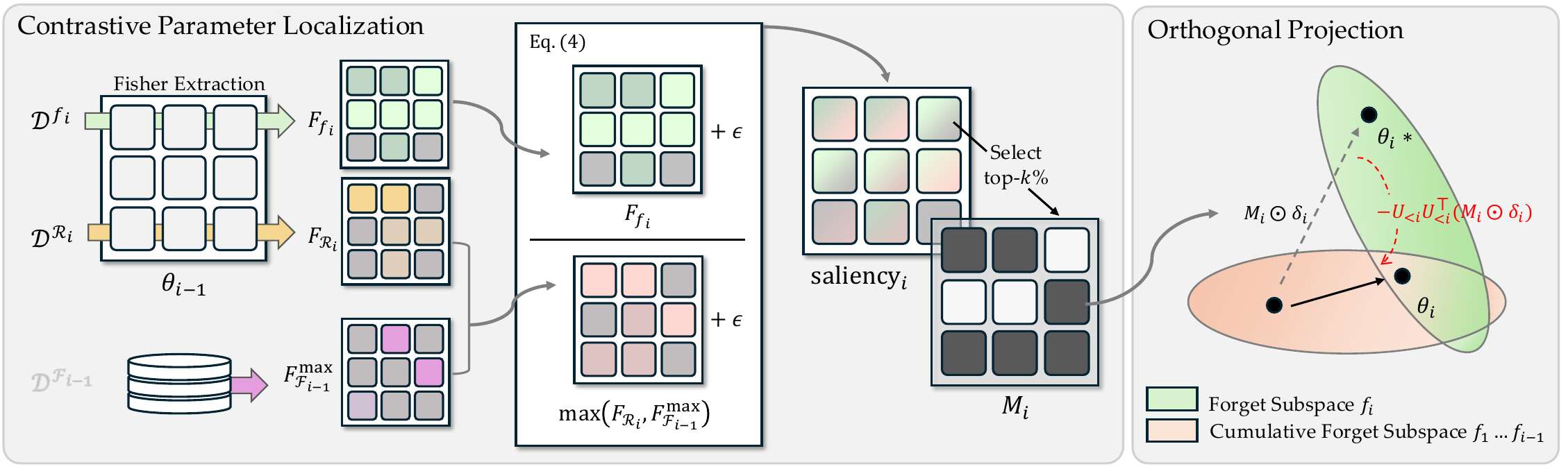}
    \caption{\textbf{Method overview.} (a) At sequence $i$, the saliency score~\eqref{eq: saliency} compares the Fisher information of the current forget set ${f_i}$ against the maximum across the $\mathcal{F}_{i-1}$. Selecting the top-$k\%$ from saliency map yields the mask $M_i$, which restricts updates to forget-relevant parameters. On (b), without intervention, the optimizer step $\delta$ would move from $\theta_{i-1}$ to a configuration outside the orange region (dashed), reverting prior unlearning. We project $\delta$ onto the orthogonal complement of $U_{<i}$,
    so the resulting $\theta_i$ remains in the intersection of both regions: $f_i$ is unlearned without disturbing $\mathcal{F}_{i-1}$.
}
    \label{fig3:method}
\end{figure}

Motivated by the above, we propose CORTIS, a \textbf{C}umulative \textbf{ORT}hogonal \textbf{I}dentity \textbf{S}uppression.
Figure \ref{fig3:method} illustrates two-step approaches of CORTIS. 
To prevent catastrophic re-learning, we constrain each unlearning sequence from interfering with prior unlearning sequences via orthogonal projection of the weight updates against an accumulated subspace.
Applied to the full model, however, this projection is inefficient: it acts on directions throughout the entire parameter space, including regions unrelated to speaker identity, and unnecessarily restricts updates that pose no risk of reversion. 
We therefore first localize the update to a small subset of parameters most relevant to forgetting the current speaker, and apply the projection only within this subset. 
This minimizes interference with retain quality and the protective effect on prior speakers.

\subsection{Contrastive Parameter Localization}\label{subsec:mask}

State-of-the-art ZS-TTS models exhibit highly entangled representations, as their components are not explicitly modularized to separate speaker identity from content generation and other speech attributes. 
Continual unlearning is particularly vulnerable in this setting.
Without constraint, the optimizer can update parameters unrelated to speaker identity, and more critically, can overwrite parameters that drove successful unlearning on previous speakers.
We therefore restrict each update to the forget-relevant subset of parameters to localize updates. 
This serves a few purposes: it guards interference with previously unlearned speakers, protects current speaker against being overwritten in future updates, and concentrates each unlearning sequence efficiently on most relevant parameters.

For each unlearning sequence $i$, we compute the diagonal Fisher Information Matrix $F_{f_i}$ of the forget loss on speaker set $f_i$'s data, and construct a saliency map that is high where forgetting is important and low where retention or prior unlearning is important:
\begin{equation}
\label{eq: saliency}
\text{saliency}_i = \frac{F_{f_i} + \epsilon}{\max\bigl(F_{\mathcal{R}_{i}},\, F_{f_1},\, \dots,\, F_{f_{i-1}}\bigr) + \epsilon},
\end{equation}
where $F_{\mathcal{R}_i}$ is the Fisher Information Matrix computed over the remain data.
In practice, $F_{\mathcal{R}_i}$ does not necessarily have to be obtained from full remain set $\mathcal{R}_i$ at each request; a fixed subset suffices and can be reused across sequences.
The element-wise max in the denominator acts as the soft guard: any parameter important for retain quality or for any prior forget speaker is pushed toward the bottom of the saliency ranking. The top-$k\%$ of $\text{saliency}_i$ globally defines the trainable mask $M_i$; remaining parameters are frozen during updates at sequence $i$.

\subsection{Orthogonal Projection on Cumulative Forget Subspace}\label{subsec:subspace}

Parameter masking concentrates each unlearning step on a forget-relevant region, sustaining prior unlearning at the parameter level while keeping fine-tuning tractable at the scale of modern ZS-TTS models. 
However, masking does not by itself constrain how parameters within that region are updated. 

After completing the unlearning run for speaker $f_i$, we extract a rank-$R$ orthonormal basis $U_i$ summarizing the directions the optimizer used during that run, following gradient-projection memory recipe of OGD~\citep{farajtabar2020ogd} and GPM~\citep{saha2021gradient}. 
We collect gradient snapshots at fixed intervals throughout training, stack them, and compute a truncated SVD; the top-$R$ left singular vectors with their singular values $\Sigma_i$ form the per-speaker artifact. 
To ensure each new basis captures only directions not already covered by prior speakers, we subtract the prior subspace from each gradient snapshot before SVD, so $U_i$ is orthogonal to $U_1, \dots, U_{i-1}$ by construction.

A direct accumulation strategy would form $[U_1 \mid \cdots \mid U_i]$ and orthonormalize. 
This grows the cumulative basis as $\sum_k r_k$ across speakers, so per-step projection cost increases with the request sequence length. 
In a deployed ZS-TTS system, the number of unlearning requests has no fixed horizon, and unbounded growth of the prior subspace is impractical. 
We therefore maintain a fixed-rank merged basis: at sequence $i+1$, we form the energy-weighted column stack $\Phi_i = [U_1\Sigma_1 \mid \cdots \mid U_i\Sigma_i]$, in which each column is scaled by its singular value to encode how much that direction was used during the corresponding speaker's run, and take the rank-$R_{\text{merge}}$ truncated SVD of $\Phi_i$. 
The resulting $U_{<i}$ has orthonormal columns by construction and bounds projection cost at a constant $R_{\text{merge}}$ regardless of $i$. 
After every optimizer step, the weight delta $\delta$ in the trainable mask is projected onto the orthogonal complement of $U_{<i}$:
\begin{equation}
\delta \leftarrow \delta - U_{<i} U_{<i}^\top \delta.
\end{equation}

\section{Experiments}
\label{sec:experiments}





\subsection{Evaluation Scenario}
\label{subsec:datasets}


We evaluate \textsc{CORTIS} under the sequential unlearning scenario formalized in Section~\ref{subsec:problem}. At each unlearning sequence $i$, the model receives a request to unlearn a speaker $f_i$, and only data from $f_i$ is accessible; data for previously forgotten speakers $\mathcal{F}_{i-1}$ is no longer available.
At each sequence, the model is initialized from the previous sequence, $\theta_{i-1}$.

\textbf{Backbone.} All methods are applied to VoiceBox~\citep{le2023voicebox}, a flow-matching ZS-TTS model with 24 transformer layers. The pre-trained checkpoint is identical across all baselines and our method to ensure fair comparison. Further training and inference details are provided in Appendix~\ref{appx:backbone_implementation}.

\textbf{Datasets.} We use LibriHeavy~\citep{kang2024libriheavy} as the pre-training corpus. Following prior work~\cite{donotmimic}, we utilize their filtered set of forget speakers, each with approximately 20 minutes of speech audio. 
We randomize the request order to select forget speaker and sequences, using the LibriHeavy speaker IDs in order of $1166 \rightarrow 7199 \rightarrow 3912 \rightarrow 9437 \rightarrow 8866$.
Remain set evaluation is conducted on LibriSpeech test-clean~\citep{panayotov2015librispeech}, following ZS-TTS protocols~\citep{le2023voicebox}. For details on forget set, please refer to Appendix~\ref{appx:forget_spks}.

\subsection{Baselines}
\label{subsec:baselines}

We compare \textsc{CORTIS} against four baselines, spanning speaker-identity unlearning methods and continual unlearning methods adapted from the vision domain. Specific implementation details on baselines are elaborated in Appendix~\ref{appx:baseline_implementation}.

\textbf{Speaker-identity unlearning baselines.} 
We adopt baselines shown effective for ZS-TTS domain \citep{donotmimic}. \textbf{Sample-Guided Unlearning (SGU)} concatenates a forget-speaker utterance with a remain-speaker utterance and trains the model to predict the masked remain-speaker region, using a sample from the retain set. \textbf{Teacher-Guided Unlearning (TGU)} instead generates a text-aligned random-voice target from the pre-trained teacher conditioned only on the transcript $y$, and trains the student to match this target when conditioned on the forget speaker's prompt. Both methods were originally proposed for joint (single-step) unlearning of a fixed forget set.

\textbf{Continual unlearning baselines.} We additionally adapt two regularizers explored in prior continual unlearning works~\citep{lee2025continual}. \textbf{Update Normalization (UN)} penalizes the norm of the parameter update relative to the previous checkpoint, $\|\theta_i - \theta_{i-1}\|$, to limit cumulative parameter drift across sequential requests. \textbf{Selective Fine-tuning (SelFT)} restricts updates at each step to the top-$k\%$ most important parameters, where importance is computed from gradient magnitudes on the forget data. We compose both regularizers with the TGU  
to isolate the effect of the regularization mechanism. For other continual unlearning baselines that violate RTBF, please refer to Appendix \ref{appx:rtbfx_baselines}.

\subsection{Evaluation Metrics}
\label{subsec:metrics}

We evaluate with two quantitative metrics. \textbf{Word Error Rate (WER)} assesses transcription fidelity of the generated speech. Audio synthesized by $\theta_i$ is transcribed with a HuBERT-Large ASR model~\citep{hsu2021hubert} fine-tuned on LibriSpeech, and compared to the ground-truth transcript. 
We report \textbf{W-R} as WER on the remain set: LibriSpeech test-clean. \textbf{W-F} is WER averaged over all forget sets $\{f_1, \dots, f_i\}$. \textbf{Speaker Similarity (SIM)} measures the cosine similarity between WavLM-TDCNN~\citep{chen2022wavlm} speaker embeddings of the prompt audio and the generated audio. \textbf{S-R} on LibriSpeech test-clean quantifies retention of the model's zero-shot voice-cloning ability on unseen speakers. For the forget set, we report SIM per forget speaker: \textbf{S-$f_i$} denotes the mean SIM over evaluation samples of forget speaker $f_i$, computed using $f_i$'s own utterances as prompts. Each sample is reported on mean over 3 generation seeds. To interpret these SIM values, we calibrate retention and forgetting thresholds against the real-world speaker similarity distribution; the resulting bounds (S-R$<0.46$ a retention failure, S-F$>0.32$ a forgetting failure) are used throughout Table~\ref{tab:main_results} and derived in Appendix~\ref{appx:speaker_distribution}.

\subsection{Experimental Results}\label{subsec:results}

\begin{table}[t]
\caption{Continual unlearning results for sequence ($f_1 \!\rightarrow\! f_2 \!\rightarrow\! f_3$). 
Each group reports metrics evaluated after the corresponding unlearn request. 
S-$f_i$ denotes speaker similarity for forget speaker $i$; for Request 2, the previously unlearned speaker's SIM is reported to measure persistence of unlearning. The Original model values reflect the pretrained model's baseline. \textcolor{failcolor}{Orange} marks successful retention preservation (S-R$\geq 0.46$) and \textcolor{shadecolor}{green} marks successful forgetting (S-$f_i < 0.32$).}
\vskip 0.1in
\label{tab:main_results}
\centering
\resizebox{\linewidth}{!}{
\begin{tabular}{l cccc ccccc cccccc}
\toprule
& \multicolumn{4}{c}{After Request 1} 
& \multicolumn{5}{c}{After Request 2}
& \multicolumn{6}{c}{After Request 3}\\
\cmidrule(lr){2-5} \cmidrule(lr){6-10} \cmidrule(lr){11-16}
Method 
& W-R$\downarrow$ & W-F$\downarrow$ & S-R$\uparrow$ & S-$f_1$$\downarrow$ 
& W-R$\downarrow$ & W-F$\downarrow$ & S-R$\uparrow$ & S-$f_1$$\downarrow$ & S-$f_2$$\downarrow$
& W-R$\downarrow$ & W-F$\downarrow$ &  S-R$\uparrow$ & S-$f_1$$\downarrow$ & S-$f_2$$\downarrow$ & S-$f_3$$\downarrow$ \\
\midrule
Original
& 2.1 & 2.6 & 0.649 & 0.721
& 2.1 & 2.5 & 0.649 & 0.721 & 0.674 
& 2.1 & 2.5 & 0.649 & 0.721 &  0.674 & 0.866 \\
\midrule
SGU & 2.7 & 2.5 & {\color{failcolor}0.479} & {\color{shadecolor}0.165}
& 2.8 & 2.6 & 0.348 & {\color{shadecolor}0.178} & {\color{shadecolor}0.075}
& 2.7 & 2.2 & 0.315 & {\color{shadecolor}0.233} & {\color{shadecolor}0.101} & {\color{shadecolor}0.192} \\
TGU & 2.3 & 2.5 & {\color{failcolor}0.624} & {\color{shadecolor}0.164}
& 2.5 & 3.0 & {\color{failcolor}0.563} & 0.612 & {\color{shadecolor}0.198}
& 3.0 & 2.6 & {\color{failcolor}0.582} & 0.603 & 0.546 & {\color{shadecolor}0.180} \\
\midrule
UN & 2.8 & 2.6 & {\color{failcolor}0.565} & {\color{shadecolor}0.229}
& 2.8 & 2.7 & {\color{failcolor}0.545} & 0.344 & {\color{shadecolor}0.140}
& 3.0 & 2.5 & {\color{failcolor}0.580} & 0.638 & 0.555 & {\color{shadecolor}0.106} \\
SelFT & 2.7 & 2.5 & {\color{failcolor}0.592} & {\color{shadecolor}0.154}
& 2.8 & 2.6 & {\color{failcolor}0.585} & 0.482 & {\color{shadecolor}0.077}
& 2.7 & 2.3 & {\color{failcolor}0.548} & 0.553 & 0.434 & {\color{shadecolor}0.110} \\
\midrule
\textbf{CORTIS}  
& 2.9 & 2.6 & {\color{failcolor}0.602} & {\color{shadecolor}0.162}
& 2.9 & 2.6 & {\color{failcolor}0.553} & {\color{shadecolor}0.185} & {\color{shadecolor}0.122}
& 2.8 & 2.6 & {\color{failcolor}0.557} & {\color{shadecolor}0.172} & {\color{shadecolor}0.148} & {\color{shadecolor}0.124} \\
\bottomrule
\end{tabular}}
\end{table}

Table~\ref{tab:main_results} reports continual unlearning results across three sequential requests. 
\textsc{CORTIS} is the only method that keeps every previously forgotten speaker forgotten across the full sequence while retaining competitive remain-set quality: after Request 3, all three forget-speaker similarities remain below $0.18$ (S-$f_1{=}0.172$, S-$f_2{=}0.148$, S-$f_3{=}0.124$), while S-R holds at $0.557$. Every baseline collapses on at least one of these axes.

\textbf{Are prior speaker identity unlearning methods capable of continual unlearning?} TGU achieves strong unlearning at Request 1 (S-$f_1{=}0.164$), but the previously unlearned speaker reverts sharply once a new request is processed: S-$f_1$ jumps to $0.612$ after Request 2 and remains at $0.603$ after Request 3, indicating that the identity of $f_1$ has been effectively re-learned by the model. The same pattern recurs for $f_2$ (S-$f_2{=}0.546$ at Request 3). SGU avoids this reversion. All forget-speaker similarities stay below $0.24$, but the retain-set quality degrades monotonically across requests, with S-R falling from $0.479$ to $0.315$ over the sequence. In other words, neither speaker-identity unlearning baseline is stable: TGU exhibits catastrophic re-learning of forgotten speakers, while SGU exhibits catastrophic forgetting of retain-set capability.

\textbf{Can't we simply apply prior continual unlearning methods?} The two regularizing methods from image domain~\citep{lee2025continual} succeed at limiting drift on the retain set. UN holds S-R at $0.580$ and SelFT at $0.548$ after Request 3, but neither prevents the forgotten speakers from re-emerging. UN's S-$f_1$ rises to $0.638$ at Request 3 (from $0.229$ at Request 1), and SelFT's reaches $0.553$. This isolates the failure: limiting parameter drift \begin{wrapfigure}{r}{0.43\linewidth}
    \vspace{-0.4cm}
    \centering
    \includegraphics[width=\linewidth]{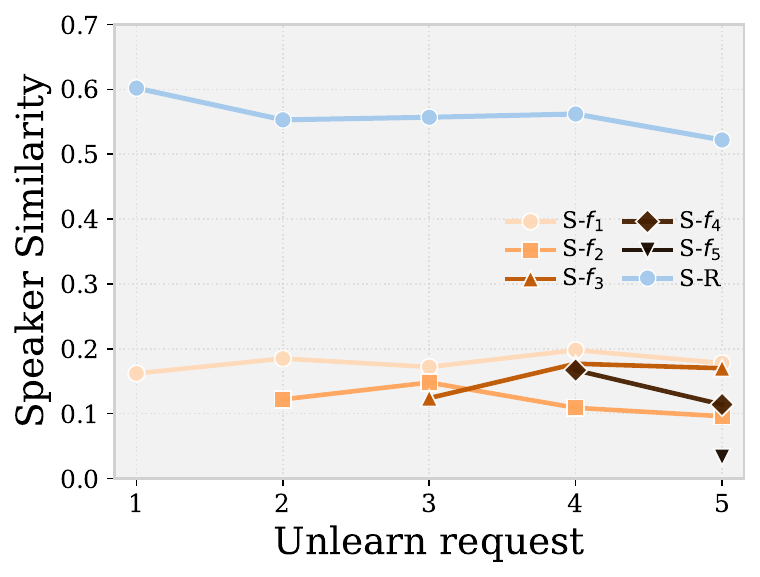}
    \vspace{-.5cm}
    \caption{CORTIS across 5 sequential unlearning requests. 
    }
    \label{fig3:unlearn_five}
    \vspace{-.8cm}
\end{wrapfigure}
preserves remain-set performance, but is not a sufficient mechanism for protecting previously forgotten identities. 
The directions along which $f_1$'s identity is encoded are not specifically guarded once the unlearning signal switches to $f_2$, $f_3$, and so on.

\textbf{Does CORTIS scale on longer sequences?}
Figure~\ref{fig3:unlearn_five} extends \textsc{CORTIS} to a longer sequence of 5 requests.
We test whether the stability observed in Table~\ref{tab:main_results} persists as more speakers are unlearned. Three observations stand out. 
First, retain-set quality is maintained when scaled: across requests, S-R does not face a sharp collapse but rather maintains a near-flat segment. 
This indicates that CORTIS does not compound interference linearly with sequence length.
Second, the worst-case forget similarity stays bounded across the full sequence: S-$F$ across all forgotten speakers at each step, remains under $0.2$ throughout. 
The per-speaker dots show no individual speaker exceeding this bound even after additional requests have been processed. 
Third, and most importantly for privacy, the speaker forgotten at Request 1 does not leak as later requests accumulate. 
Its similarity remains within the same band as recently unlearned speakers, demonstrating that the durability of unlearning is preserved across the entire history. For specific scores at each unlearn request, please refer to Appendix \ref{appx:unlearn_5}.

\subsection{Ablation and Analysis}
We conduct ablations to isolate the contribution of components in \textsc{CORTIS}: the rank $R$ of the per-sequence orthonormal basis used for projection, and the necessity of orthogonal projection itself on top of contrastive parameter masking. Both ablations follow the same sequence setup as Table~\ref{tab:main_results}.

\begin{table}[t]
\caption{Ablation on orthogonal projection evaluated on sequence ($f_1 \!\rightarrow\! f_2 \!\rightarrow\! f_3$). Results at step 1 are omitted as projection is inactive at the first request and values are identical to those in Table \ref{tab:main_results}.
}
\vskip 0.1in
\label{tab:ablation_mask}
\centering
\resizebox{\linewidth}{!}{%
\begin{tabular}{l ccccc cccccc}
\toprule
& \multicolumn{5}{c}{After Request 2}
& \multicolumn{6}{c}{After Request 3}\\
\cmidrule(lr){2-6} \cmidrule(lr){7-12}
Variant
& W-R$\downarrow$ & W-F$\downarrow$ & S-R$\uparrow$ & S-$f_1$$\downarrow$ & S-$f_2$$\downarrow$
& W-R$\downarrow$ & W-F$\downarrow$ & S-R$\uparrow$ & S-$f_1$$\downarrow$ & S-$f_2$$\downarrow$ & S-$f_3$$\downarrow$\\
\midrule

w/o Projection& 2.7 & 2.4 & 0.528 & 0.200 & 0.103
& 2.8 & 2.3 & 0.546 & 0.334 & 0.397 & 0.120\\
\midrule
w. Projection
& 2.9 & 2.6 & 0.553 & 0.185 & 0.122 
& 2.8 & 2.6 & 0.557 & 0.172 & 0.148 & 0.124\\
\bottomrule
\end{tabular}%
}
\end{table}

\begin{table}[t]
\caption{Ablation on the mask $M_i$. Evaluated on sequence ($f_1 \!\rightarrow\! f_2 \!\rightarrow\! f_3$).
}
\vskip 0.1in
\label{tab:ablation_mask}
\centering
\resizebox{\linewidth}{!}{%
\begin{tabular}{l cccc ccccc cccccc}
\toprule
& \multicolumn{4}{c}{After Request 1}
& \multicolumn{5}{c}{After Request 2}
& \multicolumn{6}{c}{After Request 3}\\
\cmidrule(lr){2-5} \cmidrule(lr){6-10} \cmidrule(lr){11-16}
Variant
& W-R$\downarrow$ & W-F$\downarrow$ & S-R$\uparrow$ & S-$f_1$$\downarrow$ 
& W-R$\downarrow$ & W-F$\downarrow$ & S-R$\uparrow$ & S-$f_1$$\downarrow$ & S-$f_2$$\downarrow$
& W-R$\downarrow$ & W-F$\downarrow$ & S-R$\uparrow$ & S-$f_1$$\downarrow$ & S-$f_2$$\downarrow$ & S-$f_3$$\downarrow$\\
\midrule
$k$=20
& 2.8 & 2.5 & 0.625 & 0.158 
& 2.7 & 2.6 & 0.601 & 0.194 & 0.112 
& 2.8 & 2.6 & 0.523 & 0.155 & 0.082 & 0.132 \\
$k$=30
& 2.9 & 2.6 & 0.602 & 0.162 
& 2.9 & 2.6 & 0.553 & 0.185 & 0.122
& 2.8 & 2.6 & 0.557 & 0.172 & 0.148 & 0.124\\
\bottomrule
\end{tabular}%
}
\end{table}

\textbf{Masking without projection.} 
To test whether masking alone is sufficient, we ablate the projection step and retain only the contrastive saliency mask $M_i$ (Table~\ref{tab:ablation_mask}). 
Masking by itself partially mitigates re-learning. It soft-excludes parameters important for prior speakers from the trainable set providing a parameter-level guard, but the effect is incomplete. 
Within the masked region, gradient updates can still move along directions previously used to forget $f_1, \dots, f_{i-1}$. S-$f_{1}$ and S-$f_{2}$ revert back to 0.334 and 0.397, respectively, at Request $3$.
This indicates that parameter-level localization and direction-level protection are complementary: masking concentrates the update on the right parameters, but only projection prevents those parameters from moving in directions that revert prior unlearning.

\textbf{Projection under varied contrastive parameter localization.}
The previous ablation removes projection while keeping the mask; here we vary the mask while keeping projection fixed, isolating the inverse direction. Table~\ref{tab:ablation_mask} reports results for $k \in \{20, 30\}$, the fraction of parameters retained as trainable by the contrastive parameter localization. A smaller mask ($k{=}20$) restricts the update to a tighter forget-relevant region; a larger mask ($k{=}30$) admits more parameters into the trainable set. Both configurations achieve comparable forget-set similarity at every request, indicating that projection successfully prevents catastrophic re-learning across a range of mask sparsities; protection of prior identities is not contingent on a particular mask budget.
The cost of a smaller mask appears on the retain side: $k{=}20$ degrades S-R from $0.557$ to $0.523$ at Request 3, reflecting that an over-restricted mask leaves the forget loss with too few parameters to make progress without inducing larger, less surgical updates within the projected subspace.
This complements the masking-only ablation: masking and projection address parameter-level and direction-level interference, respectively, and the projection mechanism is robust to the specific choice of mask budget within the regime we tested.

\textbf{Practicality.} \textsc{CORTIS} adds three sources of overhead beyond a sequential TGU run: (i) Fisher information computation on the forget speaker's data once per request, (ii) truncated SVD on collected gradient snapshots once per request, and (iii) per-step projection of weight updates against the cumulative basis. 
Table~\ref{tab:cost} reports the wall-clock cost and peak GPU memory of each component.
The Fisher pass adds approximately 30 minutes per request, the per-speaker SVD on collected gradient snapshots adds 3 seconds, and per-step projection 
\begin{wraptable}{r}{0.5\linewidth}
\vspace{-\intextsep}
\caption{Computational cost at the third sequential unlearning request ($i{=}3$). Peak GPU memory and total time reflect the cost of producing $\theta_3$ from $\theta_2$ on two NVIDIA A100 80GB.}
\vskip 0.1in
\label{tab:cost}
\centering
\small
\begin{tabular}{l ccc}
\toprule
 Method & Steps & Mem.\,(GB) & Time\,(hours)\\
\midrule
TGU (seq.)        &10K& 30.8 & 29 \\
TGU (cum.)      &30K& 30.8 & 87.5 \\
UN                &10K& 30.8 & 29 \\
SelFT             &11K& 48.5 & 22 \\ 
\midrule
\textbf{CORTIS} &3K& 49.3 & 3.5 \\
\bottomrule
\end{tabular}
\vspace{-0.3cm}
\end{wraptable}
adds approximately 0.5 s per optimizer step (\textit{i.e.}, 8.48 s $\rightarrow$ 8.76 s, with 4-step gradient accumulation), a 3.3\% overhead.
In aggregate, \textsc{CORTIS} requires 3.5 hours per unlearning request, comparable to sequential TGU (29 hours) and substantially below the cost of re-running TGU on the cumulative forget set (87.5 hours after 3 requests, scaling linearly with $i$). 
Crucially, this is the relevant comparison: re-running TGU on cumulative data is the only sequential-baseline alternative that does not catastrophically re-learn, but it (a) violates forget-data non-retention (C2), and (b) scales quadratically across the model's lifetime. \textsc{CORTIS} achieves comparable durability at constant per-request cost. For step sizes, please refer to Appendix~\ref{appx:experiment_setting}.

\section{Limitations} 
Here we outline several limitations that scope the contributions of this work and suggest natural directions for follow-on study.

\textbf{Adversarial robustness.} Our threat model assumes a service provider honestly applying unlearning upon receiving an RTBF request. We do not study adversarial scenarios in which the released model parameters are subject to fine-tuning, prompt-engineering, or activation-level attacks that attempt to recover forgotten identities. Empirical robustness of unlearned ZS-TTS models against such attacks is an important open problem and is orthogonal to the durability question we address.

\textbf{Backbone scope.} To the best of our knowledge, speaker identity unlearning has only been studied on VoiceBox~\citep{le2023voicebox, donotmimic}, and we follow this setting to enable direct comparison with prior work. The mechanisms underlying \textsc{CORTIS} (Fisher saliency, gradient subspace projection) are architecture-agnostic in principle, but cross-architecture validation on autoregressive codec-based systems such as VALL-E~\citep{valle} or diffusion-based systems such as NaturalSpeech~\citep{tan2024naturalspeech} is left to future work. 

\section{Societal Impact}

The motivation for this work is fundamentally protective. Zero-shot text-to-speech systems can clone a person's voice from seconds of reference audio, and the harms enabled by this capability disproportionately affect individuals who never consented to having their voices replicable. Continual speaker identity gives service providers a concrete procedure for honoring removal requests as they arrive, without retaining the voice data the requester asked to have erased. Unlearning durability is a necessary but not sufficient condition for deploying ZS-TTS responsibly; robust evaluation of unlearned models, transparent policies for handling RTBF requests, and adversarial robustness studies are complementary research directions that we believe should accompany continued work in this space.

\section{Conclusion}
In this paper, we systematically identify the challenges and formalize the objectives of continual speaker identity unlearning, an underexplored problem setting. Speaker identity unlearning for ZS-TTS models, by virtue of their zero-shot generalizability, are uniquely prone to reviving unlearned speakers. Even simple regularization on later forget requests is sufficient to restore previously erased identities. We demonstrate this empirically and show that unlearning methods successful in other domains fail when adopted for continual speaker identity unlearning. From this diagnosis, we identify the need for minimal interference with previous unlearning along two complementary dimensions. First, parameter-level localization through contrastive saliency masking confines each unlearning sequence to a small set of forget-specific parameters and soft-excludes parameters important to prior speakers. Second, subspace-level protection through pre-computed identity protection subspaces projects gradient updates away from directions associated with prior unlearning. Together, these mechanisms unify a stream of forget requests into a coherent unlearning task.





{
\small
\bibliographystyle{unsrt}
\bibliography{main}

@article{valle,
  title={Neural codec language models are zero-shot text to speech synthesizers},
  author={Chen, Sanyuan and Wang, Chengyi and Wu, Yu and Zhang, Ziqiang and Zhou, Long and Liu, Shujie and Chen, Zhuo and Liu, Yanqing and Wang, Huaming and Li, Jinyu and others},
  journal={IEEE Transactions on Audio, Speech and Language Processing},
  volume={33},
  pages={705--718},
  year={2025},
  publisher={IEEE}
}

@article{le2023voicebox,
  title={Voicebox: Text-guided multilingual universal speech generation at scale},
  author={Le, Matthew and Vyas, Apoorv and Shi, Bowen and Karrer, Brian and Sari, Leda and Moritz, Rashel and Williamson, Mary and Manohar, Vimal and Adi, Yossi and Mahadeokar, Jay and others},
  journal={Advances in neural information processing systems},
  volume={36},
  pages={14005--14034},
  year={2023}
}

@article{tan2024naturalspeech,
  title={Naturalspeech: End-to-end text-to-speech synthesis with human-level quality},
  author={Tan, Xu and Chen, Jiawei and Liu, Haohe and Cong, Jian and Zhang, Chen and Liu, Yanqing and Wang, Xi and Leng, Yichong and Yi, Yuanhao and He, Lei and others},
  journal={IEEE Transactions on Pattern Analysis and Machine Intelligence},
  volume={46},
  number={6},
  pages={4234--4245},
  year={2024},
  publisher={IEEE}
}

@article{gdpr,
  title={The eu general data protection regulation (gdpr)},
  author={Voigt, Paul and Von dem Bussche, Axel},
  journal={A practical guide, 1st ed., Cham: Springer International Publishing},
  volume={10},
  number={3152676},
  pages={10--5555},
  year={2017},
  publisher={Springer}
}

@article{ccpa,
  title={California consumer privacy act (CCPA)},
  author={Bonta, Rob},
  journal={Retrieved from State of California Department of Justice: https://oag. ca. gov/privacy/ccpa},
  pages={4--40},
  year={2022}
}

@inproceedings{bourtoule2021machine,
  title={Machine unlearning},
  author={Bourtoule, Lucas and Chandrasekaran, Varun and Choquette-Choo, Christopher A and Jia, Hengrui and Travers, Adelin and Zhang, Baiwu and Lie, David and Papernot, Nicolas},
  booktitle={2021 IEEE symposium on security and privacy (SP)},
  pages={141--159},
  year={2021},
  organization={IEEE}
}

@inproceedings{donotmimic,
  title={Do Not Mimic My Voice: Speaker Identity Unlearning for Zero-Shot Text-to-Speech},
  author={Kim, Taesoo and Kim, Jinju and Kim, Dong Chan and Ko, Jong Hwan and Park, Gyeong-Moon},
  booktitle={International Conference on Machine Learning},
  pages={30176--30198},
  year={2025},
  organization={PMLR}
}

@article{rtbf,
  title={The right to be forgotten},
  author={Rosen, Jeffrey},
  journal={Stan. L. Rev. Online},
  volume={64},
  pages={88},
  year={2011},
  publisher={HeinOnline}
}

@incollection{thrun1998lifelong,
  title={Lifelong learning algorithms},
  author={Thrun, Sebastian},
  booktitle={Learning to learn},
  pages={181--209},
  year={1998},
  publisher={Springer}
}

@article{parisi2019continual,
  title={Continual lifelong learning with neural networks: A review},
  author={Parisi, German I and Kemker, Ronald and Part, Jose L and Kanan, Christopher and Wermter, Stefan},
  journal={Neural networks},
  volume={113},
  pages={54--71},
  year={2019},
  publisher={Elsevier}
}

@incollection{mccloskey1989catastrophic,
  title={Catastrophic interference in connectionist networks: The sequential learning problem},
  author={McCloskey, Michael and Cohen, Neal J},
  booktitle={Psychology of learning and motivation},
  volume={24},
  pages={109--165},
  year={1989},
  publisher={Elsevier}
}

@article{kirkpatrick2017overcoming,
  title={Overcoming catastrophic forgetting in neural networks},
  author={Kirkpatrick, James and Pascanu, Razvan and Rabinowitz, Neil and Veness, Joel and Desjardins, Guillaume and Rusu, Andrei A and Milan, Kieran and Quan, John and Ramalho, Tiago and Grabska-Barwinska, Agnieszka and others},
  journal={Proceedings of the national academy of sciences},
  volume={114},
  number={13},
  pages={3521--3526},
  year={2017},
  publisher={National Academy of Sciences}
}

@inproceedings{schwarz2018progress,
  title={Progress \& compress: A scalable framework for continual learning},
  author={Schwarz, Jonathan and Czarnecki, Wojciech and Luketina, Jelena and Grabska-Barwinska, Agnieszka and Teh, Yee Whye and Pascanu, Razvan and Hadsell, Raia},
  booktitle={International conference on machine learning},
  pages={4528--4537},
  year={2018},
  organization={PMLR}
}

@inproceedings{aljundi2018memory,
  title={Memory aware synapses: Learning what (not) to forget},
  author={Aljundi, Rahaf and Babiloni, Francesca and Elhoseiny, Mohamed and Rohrbach, Marcus and Tuytelaars, Tinne},
  booktitle={Proceedings of the European conference on computer vision (ECCV)},
  pages={139--154},
  year={2018}
}

@inproceedings{zenke2017continual,
  title={Continual learning through synaptic intelligence},
  author={Zenke, Friedemann and Poole, Ben and Ganguli, Surya},
  booktitle={International conference on machine learning},
  pages={3987--3995},
  year={2017},
  organization={Pmlr}
}

@article{rolnick2019experience,
  title={Experience replay for continual learning},
  author={Rolnick, David and Ahuja, Arun and Schwarz, Jonathan and Lillicrap, Timothy and Wayne, Gregory},
  journal={Advances in neural information processing systems},
  volume={32},
  year={2019}
}

@article{shin2017continual,
  title={Continual learning with deep generative replay},
  author={Shin, Hanul and Lee, Jung Kwon and Kim, Jaehong and Kim, Jiwon},
  journal={Advances in neural information processing systems},
  volume={30},
  year={2017}
}

@article{rusu2016progressive,
  title={Progressive neural networks},
  author={Rusu, Andrei A and Rabinowitz, Neil C and Desjardins, Guillaume and Soyer, Hubert and Kirkpatrick, James and Kavukcuoglu, Koray and Pascanu, Razvan and Hadsell, Raia},
  journal={arXiv preprint arXiv:1606.04671},
  year={2016}
}

@inproceedings{mallya2018packnet,
  title={Packnet: Adding multiple tasks to a single network by iterative pruning},
  author={Mallya, Arun and Lazebnik, Svetlana},
  booktitle={Proceedings of the IEEE conference on Computer Vision and Pattern Recognition},
  pages={7765--7773},
  year={2018}
}

@inproceedings{serra2018overcoming,
  title={Overcoming catastrophic forgetting with hard attention to the task},
  author={Serra, Joan and Suris, Didac and Miron, Marius and Karatzoglou, Alexandros},
  booktitle={International conference on machine learning},
  pages={4548--4557},
  year={2018},
  organization={PMLR}
}

@inproceedings{farajtabar2020ogd,
  title={Orthogonal gradient descent for continual learning},
  author={Farajtabar, Mehrdad and Azizan, Navid and Mott, Alex and Li, Ang},
  booktitle={International conference on artificial intelligence and statistics},
  pages={3762--3773},
  year={2020},
  organization={PMLR}
}

@inproceedings{saha2021gradient,
  title={Gradient Projection Memory for Continual Learning},
  author={Saha, Gobinda and Garg, Isha and Roy, Kaushik},
  booktitle={International Conference on Learning Representations},
  year={2021}
}

@article{zeng2019continual,
  title={Continual learning of context-dependent processing in neural networks},
  author={Zeng, Guanxiong and Chen, Yang and Cui, Bo and Yu, Shan},
  journal={Nature Machine Intelligence},
  volume={1},
  number={8},
  pages={364--372},
  year={2019},
  publisher={Nature Publishing Group UK London}
}

@article{liu2025rethinking,
  title={Rethinking machine unlearning for large language models},
  author={Liu, Sijia and Yao, Yuanshun and Jia, Jinghan and Casper, Stephen and Baracaldo, Nathalie and Hase, Peter and Yao, Yuguang and Liu, Chris Yuhao and Xu, Xiaojun and Li, Hang and others},
  journal={Nature Machine Intelligence},
  volume={7},
  number={2},
  pages={181--194},
  year={2025},
  publisher={Nature Publishing Group UK London}
}

@article{cooper2024machine,
  title={Machine Unlearning Doesn't Do What You Think: Lessons for Generative AI Policy, Research, and Practice},
  author={Cooper, A Feder and Choquette-Choo, Christopher A and Bogen, Miranda and Jagielski, Matthew and Filippova, Katja and Liu, Ken and Chouldechova, Alexandra and Hayes, Jamie and Huang, Yangsibo and Mireshghallah, Niloofar and others},
  journal={Advances in neural information processing systems},
  year={2025}
}

@inproceedings{jang2023knowledge,
  title={Knowledge unlearning for mitigating privacy risks in language models},
  author={Jang, Joel and Yoon, Dongkeun and Yang, Sohee and Cha, Sungmin and Lee, Moontae and Logeswaran, Lajanugen and Seo, Minjoon},
  booktitle={Proceedings of the 61st Annual Meeting of the Association for Computational Linguistics (Volume 1: Long Papers)},
  pages={14389--14408},
  year={2023}
}

@inproceedings{gandikota2023erasing,
  title={Erasing concepts from diffusion models},
  author={Gandikota, Rohit and Materzynska, Joanna and Fiotto-Kaufman, Jaden and Bau, David},
  booktitle={Proceedings of the IEEE/CVF international conference on computer vision},
  pages={2426--2436},
  year={2023}
}

@inproceedings{fan2024salun,
  title={SalUn: Empowering Machine Unlearning via Gradient-Based Weight Saliency in Both Image Classification and Generation},
  author={Fan, Chongyu and Liu, Jiancheng and Zhang, Yihua and Wei, Dennis and Wong, Eric and Liu, Sijia},
  booktitle={International Conference on Learning Representations},
  year={2024}
}

@article{heng2023selective,
  title={Selective amnesia: A continual learning approach to forgetting in deep generative models},
  author={Heng, Alvin and Soh, Harold},
  journal={Advances in Neural Information Processing Systems},
  volume={36},
  pages={17170--17194},
  year={2023}
}

@article{lee2025continual,
  author = {Justin Lee and Zheda Mai and Jinsu Yoo and Chongyu Fan and Cheng Zhang and Wei-Lun Chao},
  title={Continual Unlearning for Text-to-Image Diffusion Models: A Regularization Perspective},
  author={Lee, Justin and Mai, Zheda and Yoo, Jinsu and Fan, Chongyu and Zhang, Cheng and Chao, Wei-Lun},
  journal={International Conference on Learning Representations},
  year={2026}
}

@article{george2025distill,
  title={Distill, Forget, Repeat: A Framework for Continual Unlearning in Text-to-Image Diffusion Models},
  author={George, Naveen and Murata, Naoki and Takida, Yuhta and Mopuri, Konda Reddy and Mitsufuji, Yuki},
  journal={arXiv preprint arXiv:2512.02657},
  year={2025}
}

@inproceedings{gao2025o3,
  title={On Large Language Model Continual Unlearning},
  author={Gao, Chongyang and Wang, Lixu and Ding, Kaize and Weng, Chenkai and Wang, Xiao and Zhu, Qi},
  booktitle={The Thirteenth International Conference on Learning Representations},
  year={2025}
}

@article{xu2026fit,
  title={FIT: Defying Catastrophic Forgetting in Continual LLM Unlearning},
  author={Xu, Xiaoyu and Du, Minxin and Fang, Kun and Liang, Zi and Xiao, Yaxin and Huang, Zhicong and Hong, Cheng and Ye, Qingqing and Hu, Haibo},
  journal={arXiv preprint arXiv:2601.21682},
  year={2026}
}

@inproceedings{wuerkaixi2025alkn,
  title={Adaptive localization of knowledge negation for continual llm unlearning},
  author={Wuerkaixi, Abudukelimu and Wang, Qizhou and Cui, Sen and Xu, Wutong and Han, Bo and Niu, Gang and Sugiyama, Masashi and Zhang, Changshui},
  booktitle={Forty-second International Conference on Machine Learning},
  year={2025}
}

@article{yadav2023ties,
  title={Ties-merging: Resolving interference when merging models},
  author={Yadav, Prateek and Tam, Derek and Choshen, Leshem and Raffel, Colin A and Bansal, Mohit},
  journal={Advances in neural information processing systems},
  volume={36},
  pages={7093--7115},
  year={2023}
}

@inproceedings{kang2024libriheavy,
  title={Libriheavy: A 50,000 hours ASR corpus with punctuation casing and context},
  author={Kang, Wei and Yang, Xiaoyu and Yao, Zengwei and Kuang, Fangjun and Yang, Yifan and Guo, Liyong and Lin, Long and Povey, Daniel},
  booktitle={ICASSP 2024-2024 IEEE International Conference on Acoustics, Speech and Signal Processing (ICASSP)},
  pages={10991--10995},
  year={2024},
  organization={IEEE}
}

@inproceedings{panayotov2015librispeech,
  title={Librispeech: an asr corpus based on public domain audio books},
  author={Panayotov, Vassil and Chen, Guoguo and Povey, Daniel and Khudanpur, Sanjeev},
  booktitle={2015 IEEE international conference on acoustics, speech and signal processing (ICASSP)},
  pages={5206--5210},
  year={2015},
  organization={IEEE}
}

@inproceedings{hsu2021hubert,
  title={HuBERT: How much can a bad teacher benefit ASR pre-training?},
  author={Hsu, Wei-Ning and Tsai, Yao-Hung Hubert and Bolte, Benjamin and Salakhutdinov, Ruslan and Mohamed, Abdelrahman},
  booktitle={ICASSP 2021-2021 IEEE International Conference on Acoustics, Speech and Signal Processing (ICASSP)},
  pages={6533--6537},
  year={2021},
  organization={IEEE}
}

@article{chen2022wavlm,
  title={Wavlm: Large-scale self-supervised pre-training for full stack speech processing},
  author={Chen, Sanyuan and Wang, Chengyi and Chen, Zhengyang and Wu, Yu and Liu, Shujie and Chen, Zhuo and Li, Jinyu and Kanda, Naoyuki and Yoshioka, Takuya and Xiao, Xiong and others},
  journal={IEEE Journal of Selected Topics in Signal Processing},
  volume={16},
  number={6},
  pages={1505--1518},
  year={2022},
  publisher={IEEE}
}

@misc{
nguyen2024unveiling,
title={Unveiling Concept Attribution in Diffusion Models},
author={Nguyen Hung-Quang and Hoang Phan and Khoa D Doan},
year={2024},
url={https://openreview.net/forum?id=kdriw2a8sl}
}

@inproceedings{kongdiffwave,
  title={DiffWave: A Versatile Diffusion Model for Audio Synthesis},
  author={Kong, Zhifeng and Ping, Wei and Huang, Jiaji and Zhao, Kexin and Catanzaro, Bryan},
  booktitle={International Conference on Learning Representations}
}

@article{vaswani2017attention,
  title={Attention is all you need},
  author={Vaswani, Ashish and Shazeer, Noam and Parmar, Niki and Uszkoreit, Jakob and Jones, Llion and Gomez, Aidan N and Kaiser, {\L}ukasz and Polosukhin, Illia},
  journal={Advances in neural information processing systems},
  volume={30},
  year={2017}
}

@article{baevski2020wav2vec,
  title={wav2vec 2.0: A framework for self-supervised learning of speech representations},
  author={Baevski, Alexei and Zhou, Yuhao and Mohamed, Abdelrahman and Auli, Michael},
  journal={Advances in neural information processing systems},
  volume={33},
  pages={12449--12460},
  year={2020}
}

@article{press2021train,
  title={Train short, test long: Attention with linear biases enables input length extrapolation},
  author={Press, Ofir and Smith, Noah A and Lewis, Mike},
  journal={arXiv preprint arXiv:2108.12409},
  year={2021}
}
}

\medskip
\newpage
\appendix
\par\noindent\rule{\textwidth}{2pt}
\begin{table}[h]
    \vspace{-0.3cm}
    \centering
    \Large
    \begin{tabular}{c}
\textbf{Appendix}
    \end{tabular}
    \vspace{-.5cm}
\end{table}
\par\noindent\rule{\textwidth}{1pt}
\tableofcontents
\addtocontents{toc}{\protect\setcounter{tocdepth}{2}}

\newpage

\section{Numerical Implementation of the Cumulative Subspace}
\label{app:cumulative_subspace_numerics}

The truncated SVD of the energy-weighted column stack $M_i = [U_1\Sigma_1 \mid \cdots \mid U_i\Sigma_i]$ described in Section~\ref{subsec:subspace} is straightforward in principle but requires care at the scale of modern ZS-TTS models. With a 30\% Fisher mask on a 24-layer transformer, the union of prior masks across $s$ speakers spans $\sim$$30$M parameters, and explicit construction of $M_i$ in single precision would occupy several gigabytes for even a handful of speakers. We use two implementation tricks to keep the merge tractable.

Rather than computing $\mathrm{SVD}(M_i)$ directly, we form the Gram matrix $G_i = M_i^\top M_i \in \mathbb{R}^{C \times C}$ where $C = \sum_{j=1}^{i} r_j$ is the total column count. For typical $r_j \leq 40$ and $i \leq 5$, $C \leq 200$, so $G_i$ fits in tens of kilobytes regardless of the parameter dimension. We then eigendecompose $G_s = V \Lambda V^\top$ and recover the top-$R_{\text{merge}}$ left singular vectors via $U_{<i} = M_i V_{:R_{\text{merge}}} \Lambda_{:R_{\text{merge}}}^{-1/2}$. The Gram matrix itself is accumulated by streaming row-chunks of $M_i$ from CPU to GPU, so peak GPU memory is bounded by the chunk size times $C$ rather than the full $|M_i|$.

\section{CORTIS Implementation}\label{appx:experiment_setting}

In this section, we provide details to reproduce our suggested method CORTIS. CORTIS is tested on top of TGU unlearning mechanism, which is further specified in Appendix~\ref{appx:baseline_implementation}.

At each unlearn sequence, CORTIS is trained with the Adam optimizer (peak learning rate $5\mathrm{e}{-5}$, $500$-step linear warm-up followed by linear decay) for initial unlearning sequence $i=1$ for 10K steps. 
Empirically we observe that when projection is not applied, unlearning takes a longer training to cleanly disentangle and unlearn a specific speaker identity.
A possible way to minimize step size at $i=1$ would be to obtain orthonormal basis from gradient snapshots of $\mathcal{D}^\mathcal{R}$, and project on the first step.
For the rest of the process in Table~\ref{tab:main_results} and Figure~\ref{fig3:unlearn_five}, we apply CORTIS with only 1K steps of training, with peak learning rate $5\mathrm{e}{-6}$, $500$-step linear warmup followed by linear decay.

For Contrastive Parameter Localization, we report the scores using $k=30$ to obtain trainable mask $M_i$ at each unlearn request. This results in freezing 70\% of model parameters at each unlearn sequence. When applying orthogonal projection on cumulative subset space, two specific steps are requires. At unlearn sequence $i$, gradients are captured every $n$ steps. Then, at unlearn sequence $i+1$, the captured gradient artifacts are used to extract a rank-$R$ orthonormal basis $U_i$. In our experiments, for initial unlearn sequence $i=1$ we collect gradients with $n=150$ (150 / 10K). For the rest, we collect gradients at $n=15$ (15/1K). For the main results in Table~\ref{tab:main_results}, we report the scores with $R=40$. The ablation on $R$ is elaborated in Table~\ref{tab:ablation_rank}.


\newpage
\section{Zero-shot Text-to-Speech Backbone Implementation}\label{appx:backbone_implementation}
In this section, we elaborate our implementation of VoiceBox~\citep{le2023voicebox} Zero-shot Text-to-Speech model used as backbone throughout our experiments.

\subsection{Backbone: VoiceBox}
VoiceBox is a non-autoregressive model for speech generation and editing that supports multilingual usage. 
VoiceBox parameterizes a continuous-time transport from a simple prior $p_0$ (typically Gaussian) to the speech distribution $p_1$ via Conditional Flow Matching (CFM), where the transformation is described by a time-dependent
flow $\phi_t$.

A neural network with parameters $\theta$ predicts the conditional vector field $v_t(w, y, x_{\text{ctx}}; \theta)$ that drives this transport. Here, $w = (1 - (1 - \sigma_{\min}) t) x_0 + t x$ is the interpolated input at time
$t$, $y$ carries frame-aligned linguistic content, $x$ is the clean speech representation (e.g., mel-spectrogram), and $x_{\text{ctx}} = (1 - m) \odot x$ is the masked acoustic context, with $m$ denoting the binary mask. 
Conditioning on $x_{\text{ctx}}$ allows the model to capture speaker and prosodic style implicitly, removing the need for explicit style labels. The trajectory of $x$ under the learned field satisfies
\begin{equation}
    \frac{d\phi_t(x)}{dt} = v_t(\phi_t(x), y, x_{\text{ctx}}); \quad \phi_0(x) = x.
\end{equation}

The model is optimized by aligning its predicted field with the ground-truth conditional field $u_t(x \mid x_1)$ that points each intermediate sample toward its target $x_1$ with CFM loss.

\subsection{Model Architecture}\label{appx:model_arch}
All baselines and our proposed method share the same backbone: VoiceBox~\citep{le2023voicebox}, configured identically across runs to isolate the effect of the unlearning algorithm. The acoustic generator is a Transformer~\citep{vaswani2017attention} augmented with U-Net-style skip connections between symmetric layers, convolutional positional embeddings~\citep{baevski2020wav2vec}, and ALiBi attention biases~\citep{press2021train}. We use 24 layers, 16 attention heads, an embedding dimension of 1024, and an FFN inner dimension of 4096.

\subsection{Duration Predictor and Vocoder}\label{appx:dur_voc}
For phoneme-level duration modeling we adopt the regression variant introduced VoiceBox~\citep{le2023voicebox}. Architecturally it mirrors the acoustic model but is considerably smaller: 8 Transformer layers, 8 heads, embedding dimension 512, and FFN dimension 2048. Training runs for 600K steps with Adam, a peak learning rate of $1\mathrm{e}{-}4$, linear warmup over the first 5K steps, and linear decay thereafter. 
Mel-spectrograms are converted back to waveforms with a Diffwave vocoder~\citep{kongdiffwave} trained on the English subset of LibriHeavy~\citep{kang2024libriheavy}.

\subsection{Pretraining}
\label{appx:pretrain}
The base VoiceBox checkpoint used as the starting point for all unlearning runs is pretrained from scratch for 500K steps~\citep{le2023voicebox}. Each mini-batch contains 75 seconds of audio. Optimization uses Adam in mixed-precision FP16, with a peak learning rate of $1\mathrm{e}{-}4$, 5K warmup steps, and linear decay over the remaining steps.

\newpage
\section{Baseline Implementations}\label{appx:baseline_implementation}
In this section, we provide implementation details for reproducibility on baselines for Table~\ref{tab:main_results}. 
All baselines share the same Zero-Shot Text-to-Speech backbone initialized from the pretrained VoiceBox~\citep{le2023voicebox}. 
Mel-spectrograms are normalized with mean -5.884 and std 2.261.
Across all methods, training uses AdamW under fp16 with gradient clipping at 0.2, EMA decay 0.9999, masking ratio sampled from [0.7, 1.0], conditioning dropout probability 0.8. Below we describe the method-specific settings.

\subsection{Sample-Guided Unlearning}
Sample-Guided Unlearning (SGU)~\citep{donotmimic} follows the original released configuration and uses a single unified batch of size 4 with gradient accumulation 4 (effective batch size 16), drawing remain and forget samples at a 0.2 forget ratio. 
Training runs for 10,000 steps with a warmup of 1,000 steps and learning rate of 1e-5.

\subsection{Teacher-Guided Unlearning}\label{appx:TGU_imp}
Teacher-Guided Unlearning (TGU)~\citep{donotmimic} follows the original released configuration. 
For each unlearn request the model is updated for 10,000 steps with 1,000 warmup steps on remain set batch size of 8, forget set batch size of 2 with gradient accumulation 4 (effective 32 / 8). Samples from the forget set are randomly selected with 20\% probability. Adam optimizer is implemented with peak learning rate of 5e-5.

\subsection{Update Norm}

Update Norm (UN)~\citep{lee2025continual}, which augments the unlearning loss with a penalty on the parameter update relative to the previous checkpoint: 
\begin{equation}
    \mathcal{L}_{\text{unlearn}}(\theta_i, f_i) + \lambda \|\theta_i - \theta_{i-1}\|_1,
\end{equation}
where $\theta_{i-1}$ is the model after the $(i{-}1)$-th request and serves as the initialization for $\theta_i$.
The baseline is implemented on top of TGU unlearning. We follow the released configurations used in the paper that introduces UN~\citep{lee2025continual} and apply $L_1$ norm and set 0.8 as penalty coefficient $\lambda$.
UN is applied as an add-on to TGU under same configurations as Section~\ref{appx:TGU_imp}.

\subsection{Selective Fine-tuning}

SelFT~\citep{lee2025continual} refers to general unlearning methods that restricts updates to the top-$k\%$ parameters most salient to the forget objective. We follow the formulation~\citep{nguyen2024unveiling} that computes parameter importance via a first-order Taylor approximation,
\begin{equation}
    \text{Importance}(d) = \big| \nabla_{\theta_{i-1}[d]} \mathcal{L}_{\text{unlearn}}(\theta_{i}, f_i) \cdot \theta_{i}[d] \big|.
\end{equation}
We select the top $30\%$ of parameters by $\text{Importance}(d)$ to form a binary gradient mask for fair comparison with CORTIS. The mask is computed once from $\theta^\dagger$ and held fixed throughout unlearning. SelFT is applied as an add-on to TGU and uses same configurations as Section~\ref{appx:TGU_imp} unless specified.

\newpage
\section{Real-world Speaker Similarities}\label{appx:speaker_distribution}

\begin{figure}[H]
    \centering
    \includegraphics[width=0.8\linewidth]{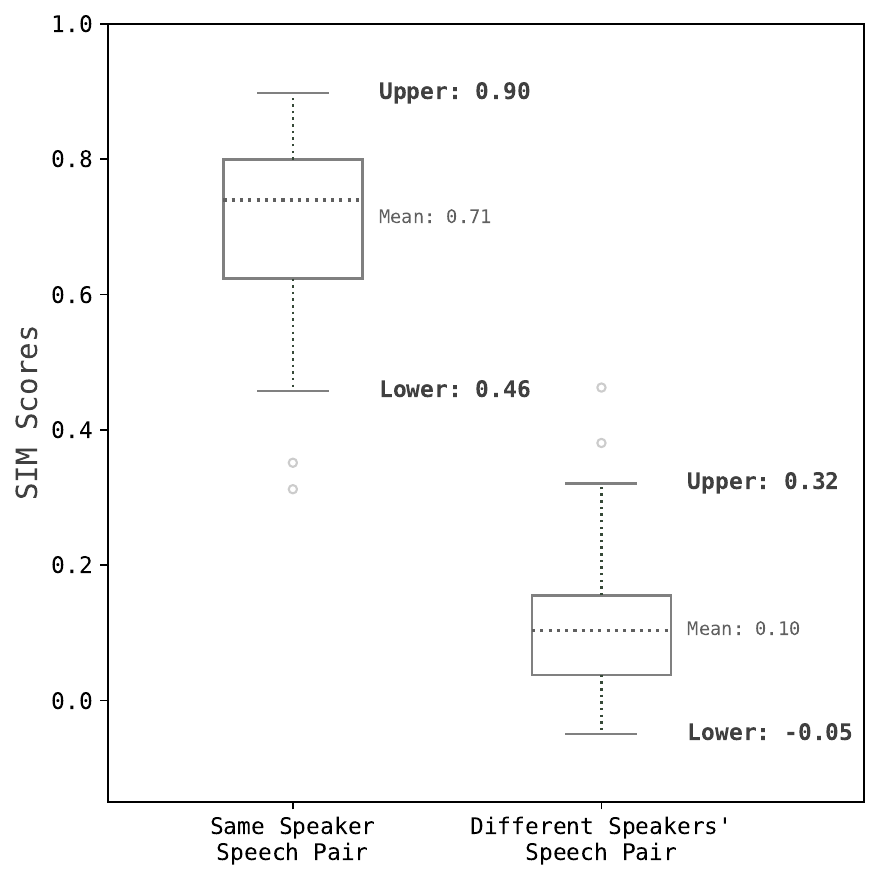}
    \caption{Empirical distribution of speaker similarity (SIM) scores across 200 randomly sampled pairs from LibriSpeech test-clean set. The upper whisker of the different-speaker distribution (0.32) and lower whisker of the same-speaker distribution (0.46) establish interpretive bounds for SIM-F and SIM-R throughout our evaluation.}
    \vskip -0.2in
    \label{fig:bound}
\end{figure}

In this section, we provide distribution of speaker similarity (SIM) on LibriSpeech test-clean dataset for an intuition on the vague boundary of: \textit{At what SIM, can we say the speakers are well retained or well forgotten?}. We visualize in Figure~\ref{fig:bound} scores obtained with randomly sampled 200 audio pairs. Evaluating the real-world bounds of different speech pairs from same speaker (left), the lower bound and upper bound are 0.46 and 0.90, respectively. For different speakers, the bounds make up a lower of -0.05 and upper of 0.32.

Based on this analysis, we set the boundary of "fail case" in retain to S-R$<0.46$, and "fail case" in forget to S-F$>0.32$. We utilize these boundaries in the main experiment in Table~\ref{tab:main_results}.

\newpage
\newpage
\section{Forget Speaker Details}\label{appx:forget_spks}

\begin{figure}[t]
    \centering
    \includegraphics[width=0.8\linewidth]{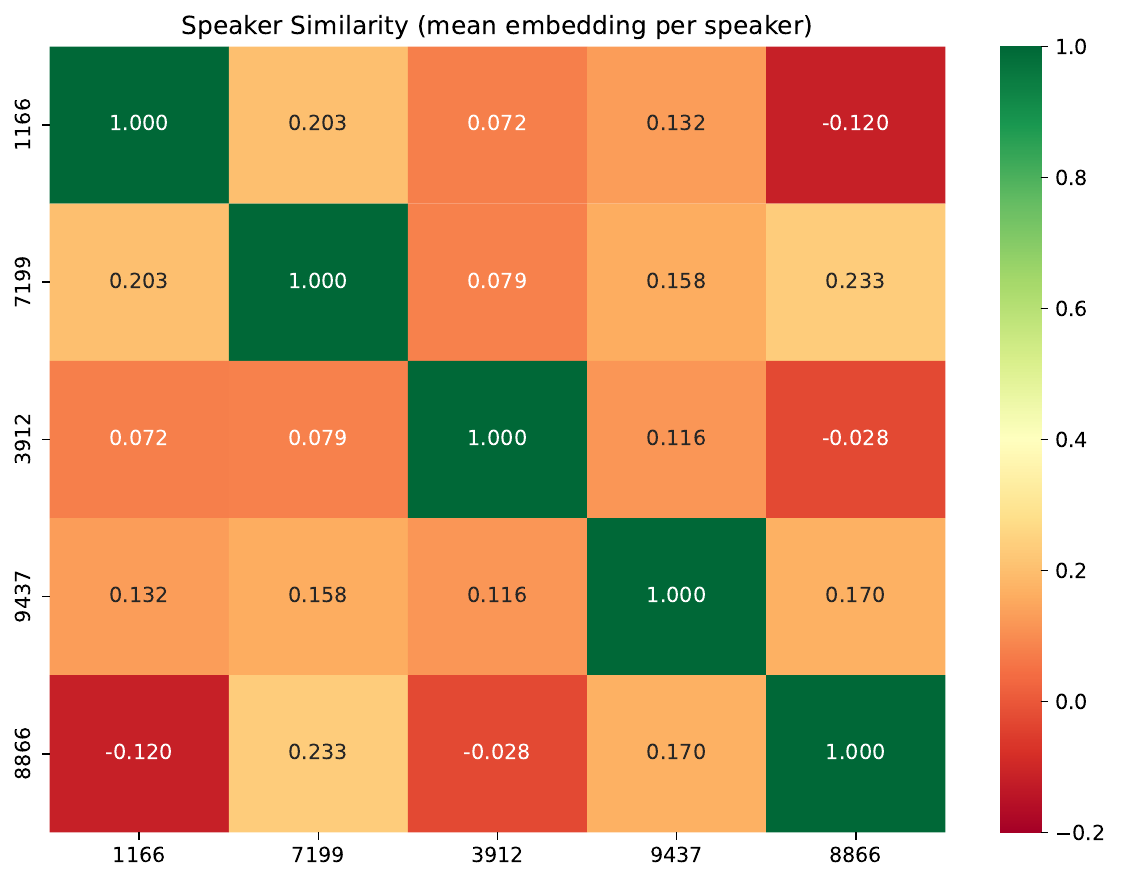}
    \caption{Pairwise cosine similarity between mean speaker embeddings for each forget speaker. Off-diagonal values are consistently low, indicating that the five speakers are well separated in embedding space.}
    \label{fig:spk_mat}
\end{figure}

To characterize the acoustic separability of the five forget speakers used in our continual unlearning sequence ($f_1{=}1166,\ f_2{=}7199,\ f_3{=}3912,\ f_4{=}9437,\ f_5{=}8866$) we examined their pairwise speaker-embedding similarity.
We sampled up to 30 utterances per speaker from the evaluation set.
For each speaker, we compute then WavLM-TDCNN~\citep{chen2022wavlm} embedding across the utterances. Embeddings were L2-normalized and pairwise cosine similarities were computed across all samples. We report cosine similarities between every pair in Figure~\ref{fig:spk_mat}.

The off-diagonal entries are uniformly low, ranging from $-0.120$ to $0.233$, with most values clustered below $0.20$. Recall that the empirical lower bound for same-speaker SIM on LibriSpeech test-clean is
$0.46$ (Appendix~\ref{appx:speaker_distribution}), so all five forget speakers fall comfortably within the different-speaker regime relative to one another. The selected sequence therefore covers a broad region of the speaker embedding space rather than a tight cluster of acoustically similar voices, ensuring that each unlearning request poses a distinct target for the model to remove.

\newpage
\section{Qualitative analysis on Contrastive Parameter Localization}
\label{appx:fisher_overlap}

A central design choice of CORTIS is the contrastive saliency in Eq.~\ref{eq: saliency}: each request's mask $M_i$ is computed by dividing the forget-set Fisher $F_{f_i}$ by the element-wise max of the remain Fisher and all prior forget Fishers, so that parameters important for retention or for any previously unlearned speaker are pushed out of the
top-$k\%$. If this mechanism works as intended, the resulting per-speaker masks should localize to largely \emph{disjoint} subsets of the model's parameters.

\begin{figure}[t]
    \centering
    \includegraphics[width=0.8\linewidth]{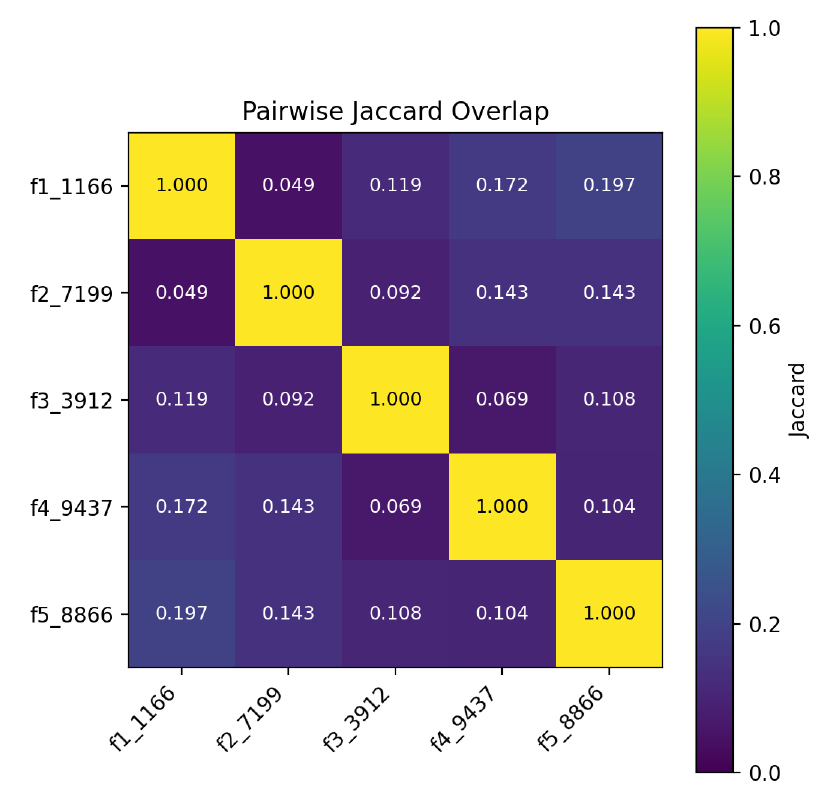}
    \caption{Pairwise Jaccard overlap between CORTIS saliency masks $M_i$
    across the five forget speakers in our continual sequence
    ($f_1{=}1166,\ f_2{=}7199,\ f_3{=}3912,\ f_4{=}9437,\ f_5{=}8866$).
    All off-diagonal values fall below $0.20$, indicating that contrastive
    saliency assigns each speaker a largely disjoint trainable subset.}
    \label{fig:fisher_overlap}
\end{figure}

We verify this empirically in Figure~\ref{fig:fisher_overlap}, which reports the pairwise Jaccard index $J(M_i, M_j) = |M_i \cap M_j|\,/\,|M_i \cup M_j|$ between the localized masks generated for the five forget speakers in our continual sequence. Off-diagonal entries are uniformly low, ranging from $0.049$ to $0.197$. This confirms that the contrastive denominator successfully decorrelates each request's mask from prior ones: forgetting $f_i$ modifies a different region of the network than forgetting $f_j$, even though all masks share the same top-$k\%$ budget. The low overlap
also supports the design intuition behind CORTIS. Because prior forget Fishers $F_{f_1}, \dots, F_{f_{i-1}}$ enter the denominator of the saliency score, the optimizer is steered away from parameters that drove earlier unlearning, protecting their effect from being overwritten.

\newpage

\section{RTBF Incompliant Baselines}\label{appx:rtbfx_baselines}

\begin{table*}[t]
\caption{Continual unlearning baselines that fail to fully meet the Right to be Forgotten (RTBF).For TIES-Merge, unlearn request 1 is not provided as they do not merge for only one speaker.}
\label{tab:appx_rtbfx_baselines}
\centering
\resizebox{\linewidth}{!}{%
\begin{tabular}{l cccc ccccc cccccc}
\toprule
& \multicolumn{4}{c}{After Request 1} 
& \multicolumn{5}{c}{After Request 2}
& \multicolumn{6}{c}{After Request 3}\\
\cmidrule(lr){2-5} \cmidrule(lr){6-10} \cmidrule(lr){11-16}
Method
& W-R$\downarrow$ & S-R$\uparrow$ & W-F$\downarrow$ & S-$f_1$$\downarrow$ 
& W-R$\downarrow$ & S-R$\uparrow$ & W-F$\downarrow$ & S-$f_1$$\downarrow$ &S-$f_2$$\downarrow$
& W-R$\downarrow$ & S-R$\uparrow$ & W-F$\downarrow$ & S-$f_1$$\downarrow$ &S-$f_2$$\downarrow$ &S-$f_3$$\downarrow$\\
\midrule
Original
& 2.1 & 0.649 & 2.6 & 0.721
& 2.1 & 0.649 & 2.5 & 0.721 & 0.674 
& 2.1 & 0.649 & 2.5 &0.721  & 0.674  &  0.866\\
\midrule
TIES-Merge
& - & - & - & -
& 2.7 & 0.558 & 2.6 & 0.241 & 0.110
& 2.7 & 0.552 & 2.8 &0.396 & 0.177 & 0.563\\
\midrule
\textbf{OURS}  
& 2.9 & 0.602 & 2.6 & 0.162
& 2.9 & 0.553 & 2.6 & 0.185 & 0.122 
& 2.8 & 0.557 & 2.6 &0.172 & 0.148 & 0.124\\


\bottomrule
\end{tabular}%
}
\vskip -0.14in
\end{table*}

In Table \ref{tab:appx_rtbfx_baselines}, we additionally compare against a model-merging baseline suggested in \citep{lee2025continual} in which each speaker is unlearned independently from the pretrained model $\theta_0$, yielding per-speaker checkpoints $\theta_1$, and the resulting weights are combined via TIES-Merging \citep{yadav2023ties}.
Concretely, for every $f \in\mathcal{F}_i$ we run the same single-speaker unlearning procedure on pre-trained model $\theta_0$ in isolation, obtaining a per-speaker unlearned checkpoint $\tilde{\theta}_f$ and the corresponding task vector $\tau_f =\tilde{\theta}f - \theta_0$. The deployed model at sequence $i$ is                       
  \begin{equation}                                                    
  \theta_i^{\text{TIES}} = \theta_0 + \lambda \cdot \tau{\mathcal{F}i},
  \end{equation}                                                        
where $\tau{\mathcal{F}_s}$ is obtained from ${\tau_f : f \in \mathcal{F}_i}$ by top-$k\%$ magnitude trimming, per-element sign election, and disjoint-mean averaging across the surviving entries ($k=20\%$, $\lambda=1.0$, following the original TIES recipe). 

This baseline violates (C2). The merge formulation requires $\theta_0$ as the additive base, so the controller must keep the pretrained model indefinitely; by Eq.~(1), $\theta_0$ already satisfies $\theta_0(x^f, y) \approx \hat{x}^{spk=f}_y$ for every $f \in \mathcal{F}i$, and therefore constitutes a fully-functional, pre-trained model held by the controller in direct violation of (C2) the underlying generative capability for every forget identity remains intact in $\theta_0$'s weights, regardless of what is suppressed in $\theta_i^{\text{TIES}}$. Extending the merge to a new request $f{i+1}$ further requires retaining either every per-speaker checkpoint ${\tilde{\theta}_f : f \in \mathcal{F}_i}$, each a deterministic function of $\mathcal{D}^f$, from which the speaker-specific direction $\tau_f$ is trivially recovered as $\tilde{\theta}_f - \theta_0$ — or the underlying datasets ${\mathcal{D}^f : f \in \mathcal{F}_i}$ themselves; both alternatives violate (C2). We therefore exclude TIES-Merge from the main table as it is not a fair comparison, nor a valid solution to the continual speaker identity unlearning problem for real world deployment of zero-shot text-to-speech models. It is also notable that OURS method performs significantly better even without having to retain model weights that violate (C2).

\newpage
\section{Scalability Across Longer Sequences}\label{appx:unlearn_5}

Table~\ref{tab:appx_unlearn_5} reports the exact numerical values plotted in
Figure~\ref{fig3:unlearn_five} (main text), provided here for completeness and
to support precise comparison across methods at each step of the unlearning
sequence.

\section{Statistical Significance}\label{appx:stats}

\begin{table}[t]
\caption{Standard deviations across evaluation speakers for the continual unlearning baselines and CORTIS results in main experiment Table~\ref{tab:main_results}.}
\vskip 0.1in
\label{tab:main_results_std}
\centering
\resizebox{\linewidth}{!}{
\begin{tabular}{l cccc ccccc cccccc}
\toprule
& \multicolumn{4}{c}{After Request 1} 
& \multicolumn{5}{c}{After Request 2}
& \multicolumn{6}{c}{After Request 3}\\
\cmidrule(lr){2-5} \cmidrule(lr){6-10} \cmidrule(lr){11-16}
Method 
& W-R & W-F & S-R & S-$f_1$ 
& W-R & W-F & S-R & S-$f_1$ & S-$f_2$
& W-R & W-F & S-R & S-$f_1$ & S-$f_2$ & S-$f_3$ \\
\midrule
UN
& 0.05 & 0.04 & 0.091 & 0.082
& 0.04 & 0.03 & 0.092 & 0.158 & 0.094
& 0.06 & 0.04 & 0.108 & 0.095 & 0.107 & 0.079 \\
SelFT
& 0.05 & 0.03 & 0.086 & 0.074
& 0.05 & 0.03 & 0.091 & 0.155 & 0.100
& 0.05 & 0.03 & 0.110 & 0.119 & 0.098 & 0.071 \\
\midrule
\textbf{CORTIS}
& 0.05 & 0.03 & 0.089 & 0.074
& 0.05 & 0.03 & 0.112 & 0.075 & 0.017
& 0.06 & 0.03 & 0.113 & 0.080 & 0.075 & 0.067 \\
\bottomrule
\end{tabular}}
\end{table}

The values in Table~\ref{tab:main_results} are computed by generating each evaluation sample under three random seeds, taking the per-sample mean, and reporting either the speaker-level (S-$f_i$) or test-set-level (W-R, S-R, W-F) average over these per-sample means. Table~\ref{tab:main_results_std} reports the corresponding standard deviations across seeds, capturing how consistent each method's behavior is.

\newpage
\section{Effect of Subspace Rank $R$}

\begin{table*}[t]
\caption{Continual unlearning results for sequence $f_1 \rightarrow f_2 \rightarrow f_3 \rightarrow f_4 \rightarrow f_5$. Here we report unlearning evaluations at $f_4$ and $f_5$ reported in Figure~\ref{fig3:unlearn_five}.
}
\vskip 0.1in
\label{tab:appx_unlearn_5}
\centering
\resizebox{\linewidth}{!}{%
\begin{tabular}{l ccccccc cccccccc}
\toprule
& \multicolumn{7}{c}{After Request 4}
& \multicolumn{8}{c}{After Request 5}\\
\cmidrule(lr){2-8} \cmidrule(lr){9-16}
& W-R$\downarrow$ & S-R$\uparrow$ & W-F$\downarrow$ & S-$f_1$$\downarrow$ &S-$f_2$$\downarrow$ &S-$f_3$$\downarrow$ &S-$f_4$$\downarrow$
& W-R$\downarrow$ & S-R$\uparrow$ & W-F$\downarrow$ & S-$f_1$$\downarrow$ & S-$f_2$$\downarrow$ &
S-$f_3$$\downarrow$ &
S-$f_4$$\downarrow$ &
S-$f_5$$\downarrow$\\
\midrule
CORTIS &
2.8 & 0.562 & 2.8 & 0.198 & 0.109 & 0.167 & 0.193 
& 2.7 & 0.527 & 2.3 &  0.178 &  0.096 & 0.170 & 0.114 & 0.033\\

\bottomrule
\end{tabular}%
}
\vskip -0.14in
\end{table*}

\begin{table}[t]
\caption{Ablation on the rank $R$ of the per-sequence orthonormal basis $U_i$ used for orthogonal projection. Evaluated on sequence ($f_1 \!\rightarrow\! f_2 \!\rightarrow\! f_3$). Results at step 1 are omitted as projection is inactive at the first request and values are identical to those in Table \ref{tab:main_results}.
}
\vskip 0.1in
\label{tab:ablation_rank}
\centering
\resizebox{\linewidth}{!}{%
\begin{tabular}{l ccccc cccccc}
\toprule
& \multicolumn{5}{c}{After Request 2}
& \multicolumn{6}{c}{After Request 3}\\
\cmidrule(lr){2-6} \cmidrule(lr){7-12}
Variant
& W-R$\downarrow$ & W-F$\downarrow$ & S-R$\uparrow$ & S-$f_1$$\downarrow$ & S-$f_2$$\downarrow$
& W-R$\downarrow$ & W-F$\downarrow$ & S-R$\uparrow$ & S-$f_1$$\downarrow$ & S-$f_2$$\downarrow$ & S-$f_3$$\downarrow$\\
\midrule
20
& 2.7 & 2.6 & 0.558 & 0.169 & 0.093
& 2.6 & 2.6 & 0.568 & 0.187 & 0.090 & 0.144\\
30
& 2.7 & 2.6 & 0.556 & 0.156 & 0.108
& 2.8 & 2.6 & 0.476 & 0.140 & 0.086 & 0.163\\
40
& 2.9 & 2.6 & 0.553 & 0.185 & 0.122 
& 2.8 & 2.6 & 0.557 & 0.172 & 0.148 & 0.124\\
\bottomrule
\end{tabular}%
}
\end{table}

The rank $R$ of each per-sequence basis $U_i$ determines how much of the gradient subspace from sequence $i$ is preserved as a constraint on subsequent updates. A small $R$ may leave the protected subspace under-specified, allowing future updates to drift back into directions that revert speaker $f_i$; a large $R$ over-constrains later sequences by reserving too many directions, restricting plasticity and degrading both forget effectiveness on $f_{i+1}, \dots$ and retain quality. Table~\ref{tab:ablation_rank} reports performance under varied $R \in \{20, 30, 40\}$. 





\end{document}